\documentclass[12pt]{article}
\usepackage{amsmath,amssymb}
\usepackage{amsthm} 
\usepackage{latexsym}
\newtheorem{thm}{Theorem}[section]
\newtheorem{prp}[thm]{Proposition}
\newtheorem{lem}[thm]{Lemma}
\newtheorem{cor}[thm]{Corollary}

\newcommand{\bpf}{\noindent {\bf Proof}. $\mbox{}$}
\newcommand{\epf}{\hfill \hbox{\raisebox{.5ex}{\fbox}$\phantom{.}$}}

\newcommand{\beq}{\begin{eqnarray}}
\newcommand{\eeq}{\end{eqnarray}}
\newcommand{\beqn}{\begin{eqnarray*}}
\newcommand{\eeqn}{\end{eqnarray*}}

\begin{document}
\title{\bf The Facets of the Bases Polytope of a Matroid and Two Consequences}
\date{}
\maketitle
\begin{center}
\author{
{\bf Brahim Chaourar} \\
{ Department of Mathematics and Statistics,\\Al Imam Mohamed Ibn Saud University (IMSIU) \\P.O. Box
90950, Riyadh 11623,  Saudi Arabia }\\{Correspondence address: P. O. Box 287574, Riyadh 11323, Saudi Arabia}}
\end{center}



\begin{abstract}
\noindent Let $M$ to be a matroid defined on a finite set $E$ and $L\subset  E$. $L$ is locked in $M$ if $M|L$ and $M^*|(E\backslash L)$ are 2-connected, and $min\{r(L), r^*(E\backslash L)\} \geq 2$. In this paper, we prove that the nontrivial facets of the bases polytope of $M$ are described by the locked subsets. We deduce that finding the maximum--weight basis of $M$ is a polynomial time problem for matroids with a polynomial number of locked subsets. This class of matroids is closed under 2-sums and contains the class of uniform matroids, the V\'amos matroid and all the excluded minors of 2-sums of uniform matroids. We deduce also a matroid oracle for testing uniformity of matroids after one call of this oracle.

\end{abstract}


\section{Introduction}

Sets and their characterisitic vectors will not be distinguished. We refer to Oxley \cite{Oxley 1992} and Schrijver \cite{Schrijver 1986} about, respectively, matroids and polyhedra terminolgy and facts.

\noindent Let $M$ to be a matroid defined on a finite set $E$. $\mathcal I(M)$,  $\mathcal B(M)$ and the function $r$ are, respectively, the class of independent sets, bases and the rank function of $M$. $M^*$, $\mathcal B^*(M)$ and the function $r^*$ are, respectively, the dual matroid, the class of cobases and the dual rank function of $M$.
For any $X\subset E$, $\mathcal B(X)$ and $\mathcal B^*(X)$  are, respectively the class of bases of $M|X$and cobases of $M^*|X$. The polyhedra $Q(M)$ and $P(M)$ are, respectively, the convex hulls of the independent sets and the bases of $M$.
Suppose that $M$ (and $M^*$) is 2-connected. A subset $L\subset E$ is called a locked subset of $M$ if $M|L$ and $M^*|(E\backslash L)$ are 2-connected, and their corresponding ranks are at least 2, i.e., $min\{r(L), r^*(E\backslash L)\} \geq 2$. It is not difficult to see that if $L$ is locked then both $L$ and $E\backslash L$ are closed, respectively, in $M$ and $M^*$ (That is why we call them locked). We denote by $\mathcal{L}(M)$ and $\ell(M)$, respectively, the class of locked subsets of $M$ and its cardinality, which is called the locked number of $M$. Given a positive integer $k$ ($k$ does not depend on $M$ or $|E|$), we say that $M$ is $k$-locked if $\ell(M) \in O(|E|^k)$.  $\mathcal L_k$ is the class of $k$-locked matroids. $M$ is 0-locked if $\mathcal L(M) = \varnothing$, i.e., $\ell(M)$ = 0 and the class of such matroids is $\mathcal L_0$. For a given nonegative integer $k$, $\mathcal L_k$ is called also a polynomially locked class of matroids. It is not difficult to see that the class of lockeds subsets of a matroid $M$ is the union of lockeds subsets of the 2-connected components of $M$. The locked structure of $M$ is the quadruple ($\mathcal P(M)$, $\mathcal S(M)$, $\mathcal L(M)$, $\rho$ ), where  $\mathcal P(M)$ and $\mathcal S(M)$ are, respectively, the class of parallel and coparallel closures, and $\rho$ is the rank function restricted to $\mathcal P(M)\cup \mathcal S(M)\cup \mathcal L(M)\cup \{\varnothing , E\}$.
\newline Given a weight function $c\in R^E$, the maximum-–weight basis problem (MWBP) is the following optimization problem:	\newline \centerline{Maximize\{$c(B)$ such that $B\in \mathcal B(M)$\}}
The corresponding maximum-–weight independent problem is clearly (polynomially time) equivalent to MWBP.
MWBP is polynomial on $|E|$ and $\theta$, where $\theta$ is the complexity of the used matroid oracle \cite{Edmonds 1971}. Even if we use the approach introduced by Mayhew \cite{Mayhew 2008} by giving the list of bases (for example) in the input, MWBP is polynomial on the size of the input. However, as Robinson and Welsh \cite{Robinson and Welsh 1980} note, no matter which of the ways to specify a matroid, the size of the input for a matroid problem on an $n$-element set is $O(2^n)$. It follows that MWBP is not polynomial in its strict sense, that is on $|E|$. We prove that MWBP is polynomial on $|E|$ for polynomially locked classes of matroids, i.e., for any matroid $M\in \mathcal L_k$ (for a fixed $k$). This class of polynomially locked matroids is closed under 2-sums and contains the class of uniform matroids, the V\'amos matroid and all the excluded minors of 2-sums of uniform matroids. These excluded minors are $M(K_4)$, $W^3$, $Q_6$ and $P_6$ \cite{Chaourar 2011}. It follows that this class is larger than 2-sums of uniform matroids.
\newline Testing Uniformity of matroids (TUM) is to provide an algorithm in which the matroid is represented by an oracle and which decides whether the given matroid is uniform or not after a number of calls on the oracle which is bounded by a polynomial in the size of the ground set. Jensen and Korte \cite{Jensen and Korte 1982} proved that there exists no such algorithm in which the matroid is represented by an independence test oracle (or an oracle polynomially related to an independence test oracle). In this paper, we give a matroid oracle which answers this question. 
\newline The remainder of the paper is organized as follows: in section 2, we give all facets of the bases polytope, then, in section 3, we deduce two consequences of this characterization. The first one is that MWBP is polynomial (time) for polynomially locked matroids, and the second one is a polynomial time algorithm via a new matroid oracle for testing if a given matroid is uniform or not. In section 4, we describe some polynomially locked classes of matroids, and we conclude in section 5.

\section{Facets of the bases polytope}

A description of $Q(M)$ was given by Edmonds \cite{Edmonds 1971} as follows.
\begin{thm} $Q(M)$ \rm{is the set of all} $x\in R^E$ \rm{such that}
$$ x(e) \geq 0    \> \mathrm{for\> any} \>  e\in E             \eqno (1) $$						
$$ x(A) \leq r(A) \> \mathrm{for\> any} \> A\subseteq E	\eqno 	(2) $$  
\end{thm}
Later, a minimal description of $Q(M)$ was given also by Edmonds \cite{Giles 1975} as follows.
\begin{thm} \rm{The inequality (2) is a facet of} $Q(M)$ \rm{if and only if} $A$ \rm{is closed and 2-connected}.
\end{thm}
It is not difficult to see that $P(M)$ is the set of all $x\in R^E$ satisfying the inequalities (1), (2) and
$$ x(E) = r(E)	\eqno	(3)$$
It seems natural to think that the inequality (2) is a facet of $P(M)$ if and only if $A$ is closed and 2-connected. This is not true because:
\begin{lem} \rm{If the inequality (2) is a facet of} $P(M)$ then $A$ is a locked subset of $M$.
\end{lem}
\bpf It suffices to prove that if $X$ is closed and 2-connected but $E\backslash L$ is not 2-connected in the dual then the inequality (2) is not a facet. In fact, there exist $A$ and $B$ two disjoint subsets of $E$ such that $E\backslash X = A\cup B$ and $r^*(E\backslash X) = r^*(A)+r^*(B)$, that is, $|E\backslash X|-–r(E)+r(X) = |A|-–r(E)+r(E\backslash A)+|B|-–r(E)+r(E\backslash B)$. It follows that: $r(E)+r(X) = r(E\backslash A)+r(E\backslash B) \geq x(E\backslash A)+x(E\backslash B) = x(E)+x(X)$, which implies the inequality (2). So the inequality (2) is redundant and cannot be a facet. \epf
\newline We give now a minimal description of $P(M)$. A part of the proof is inspired from a proof given by Pulleyblank \cite{Pulleyblank 1989} to describe the nontrivial facets of $Q(M)$. Independently, Fujishige \cite{Fujishige 1984}, and Feichtner and Sturmfels \cite{Feichtner and Sturmfels 2005}, gave a characterization of nontrivial facets of $P(M)$. We give here below a new and complete formulation with a new proof.
\begin{thm} \rm{A minimal description of} $P(M)$ is the set of all $x\in R^E$ satisfying the equality (3) and the following inequalities:
$$	x(P) \leq 1 \> \mathrm{for\>any\> parallel\> closure}\> P\subseteq E			\eqno		(4)$$
$$	x(S) \geq |S|-–1  \> \mathrm{for\> any\> coparallel\> closure}\> S\subseteq E		\eqno		(5)$$
$$	x(L) \leq r(L)	 \> \mathrm{for\> any\> locked\> subset}\> L\subseteq E			\eqno		(6)$$
\end{thm}
\bpf
Without loss of generality, we can suppose that $M$ is without parallel or coparallel elements so the inequalities (5) become as (1) and the inequalities (4) become as follows:
$$	x(e) \leq 1 \>	\mathrm{for\> any}\> e\in E		\eqno	(7)$$
Let $C(M)$ be the cone generated by the incidence vectors of the bases of $M$. It suffices to prove that the minimal description of $C(M)$ is given by (1) and the following inequalities:
$$	x(e) \leq x(E)/r(E)\>	\mathrm{for\> any}\> e\in E	\eqno	(8)$$
$$	x(L)/r(L) \leq x(E)/r(E)	\>\mathrm{for\> any\> locked\> subset}\> L \subseteq E	\eqno	(9)$$
It is not difficult to see via induction and operations of deletion and contraction that the inequalities (1) and (8) are facets of $C(M)$. It remains to prove that the inequality (9) is a facet of $C(M)$ if and only if $L$ is a locked subset of $M$. According to Lemma 2.3, it suffices to prove the inverse way.
Note that (9) is equivalent to the following inequality:
$$	(r(L)-–r(E)) x(L)+r(L) x(E\backslash L) \geq 0	\>\mathrm{for\> any\> locked\> subset}\> L \subseteq E	\eqno	(10)$$
Let $a x \geq 0$ be a valid inequality for $C(M)$ which is tight for all $B\in \mathcal B(L)$.
\newline \textbf{Claim 1:} $a_j = a_k$ for all $j$ and $k$ of $L$.
\newline Suppose this is not true. Let $X = \{j\in L\> \mathrm{such\> that}\> a_j\> \mathrm{takes\> minimum \>value\> over}\> L\}$, $Y = L\backslash X$ and $B\in \mathcal B(L)\cap \mathcal B(Y)$.
Since L is 2-connected in $M$, and since, by assumption, $X$ is a strict subset of $L$, then $r(X) > |B\cap X|$. Thus it exists $e\in X\backslash B$ such that $(B\cap X)\cup {e}$ is an independent set of $M$. It follows that there exists $f\in B\cap Y$ such that $\widetilde{B} = B\backslash {f})\cup {e}\in \mathcal B(L)$. But: $a(\widetilde{B}) = a(B)-–a(f)+a(e) < a(B)$, a contradiction.
\newline \textbf{Claim 2:} For any $X\subseteq E$, $B\in \mathcal B(X)$ if and only if $E\backslash B\in \mathcal B^*(E\backslash X)$.
\newline It suffices to prove one way and use duality for the other way.
\newline Let $B\in \mathcal B(X)$ then
$$ |B\cap X| = r(X) = |X|–-r^*(E)+r^*(E\backslash X)$$
$$ = |E|-–|E\backslash X|–-r^*(E)+r^*(E\backslash X) = r(E)-–|E\backslash X|+r^*(E\backslash X).$$
Thus,
$$ |(E\backslash B)\cap (E\backslash X)| = |E\backslash X|-–|B\cap (E\backslash X)| = |E\backslash X|-–|B|+|B\cap X|$$
$$ = |E\backslash X|-–|B|+r(E)-–|E\backslash X|+r^*(E\backslash X) = r^*(E\backslash X).$$
Since $E\backslash B$ is a basis in the dual, then $E\backslash B\in \mathcal B^*(E\backslash X)$.
\newline \textbf{Claim 3:} $a_j = a_k$ for all $j$ and $k$ of $E\backslash L$.
\newline Using claim 2, $E\backslash L$ being 2-connected and a similar argument on $E\backslash B$ as in claim 1, we conclude.
\newline \textbf{Claim 4:} $a x \geq 0$ is a multiple of inequality (10).
\newline By claims 1 and 3, $a x \geq 0$ becomes: $a_L x(L)+a_{E\backslash L} x(E\backslash L) \geq 0$. Thus, for $B\in \mathcal B(L)$, we have:
$$ a_L |B\cap L|+a_{E\backslash L} |B\cap (E\backslash L)| = 0,$$
that is, 
$$ a_L r(L)+a_{E\backslash L} (r(E)–r(L)) = 0.$$
But $a_L = r(L)–r(E)$ and $a_{E\backslash L} = r(L)$ is a solution of this equation, so we conclude. 
\epf
\section{MWBP and TUM}
Since the bases polytope is completly described by the locked structure of the matroid, so a natural matroid oracle follows.

The $k$-locked oracle
\newline \begin{tabular}{*6l}
Input: & a nonegative integer $k$ and a matroid $M$ defined on $E$. \\
Output:& (1) No if $\ell(M)\notin O(|E|^k)$, and\\
             & (2) ($\mathcal P(M)$, $\mathcal S(M)$, $\mathcal L(M)$, $\rho$ )  if $\ell(M)\in O(|E|^k)$.\\
 \end{tabular}
\newline{}
\newline{}
\newline Note that this oracle has time complexity $O(|E|^{k+1})$ because we need to count at most $|E|^{k+1}$ members of $\mathcal L(M)$ in order to know that $M$ is not $k$-locked, even if the memory complexity can be $O(|E|+\ell (M))$. Actually this matroid oracle permits to recognize if a given matroid is $k$-locked or not for a given nonegative integer $k$ (which does not depend on $M$ or $|E|$).
\newline The first consequence of Theorem 2.4 then follows.
\begin{cor} Given a nonegative integer $k$, a matroid $M\in \mathcal L_k$, the $k$-locked oracle to acess $M$ and a weight function $c\in R^E$. Then there exists a polynomial time algorithm on the size of $E(M)$ for solving MWBP in $M$.
\end{cor}
\bpf Let $M$ be a such matroid. Since $M\in \mathcal L_k$ then it can be described by its locked structure in the input of MWBP by using the k-locked oracle. MWBP is equivalent for optimizing on $P(M)$, which is also equivalent to separating on $P(M)$. Since the number of facets of $P(M)$ is $2|E|+\ell (M)$ then separating can be done on $O(|E|+\ell (M))$. But $M$ is polynomially locked, then $\ell(M)\in O(|E|^k)$  and separating on $P(M)$ can be done on $O(|E|^k)$.
\epf
\newline The k-locked oracle is stronger than the rank and the independence oracles for polynomially locked matroids:
\newline We can get the rank of any subset $X\subseteq E$ by choosing the weight function $c$ equal to the characteristic vector of $X$ and optimizing on $P(M)$, which can be done in polynomial time. The obtained optimum value of c is the requested rank.
\newline For the independence oracle, we can decide if a subset $X\subseteq E$ is independent or not by choosing the same previous weight function and decide that $X$ is independent if it is included in the optimum basis, and not if else.
\newline For testing uniformity of matroids, we need the following result \cite{Chaourar 2011}.
\begin{thm} Given a 3-connected matroid $M$.
\newline $M$ is uniform if and only if $\ell(M)=0$.
\end{thm}
We can see through the proof of this theorem that $\ell(M)=0$ if $M$ is uniform whatever its connectivity.
For disconnected matroids, we have the following result \cite{Oxley 1992}.
\begin{thm} Given a disconnected matroid $M$.
\newline $M$ is uniform if and only if $r(M)=|E|$ or $r(M)=0$.
\end{thm}
For 2-connected matroids, we can write:
\begin{prp} Given a 2-connected matroid $M$. $M$ is uniform if and only if one of the following properties holds:
\newline (i) $\ell (M)=0$ and $|\mathcal P(M)|=|E|=|\mathcal S(M)|$;
\newline (ii) $|\mathcal P(M)|=1$;
\newline (iii) $|\mathcal S(M)|=1$.
\end{prp}
So we can now characterize uniform matroids as follows.
\begin{cor} $M$ is uniform if and only if one of the following properties holds:
\newline (i) $\ell (M)=0$ and $|\mathcal P(M)|=|E|=|\mathcal S(M)|$;
\newline (ii) $|\mathcal P(M)|=1$;
\newline (iii) $|\mathcal S(M)|=1$;
\newline (iv) $r(M)=|E|$;
\newline (v) $r(M)=0$.
\end{cor}
A natural matroid oracle follows.

The locked number oracle
\newline \begin{tabular}{*4l}
Input:&  a matroid $M$ defined on $E$. \\
Output:&   $\ell (M)$, $r(M)$, $|\mathcal P(M)|$, $|\mathcal S(M)|$.\\
\end{tabular}
\newline {}
\newline Note that, except for $\ell (M)$, all other outputs of this oracle can be computed in a polynomial time given a locked structure of $M$.
We can now give an algorithm which tests if a given matroid is uniform after one call of the locked number oracle.

Testing Uniformity of Matroids
\newline \begin{tabular}{llllllllllllllll}
Input:&  a matroid $M$ defined on $E$. \\
Output:&   (a) $M$ is uniform if one of the following properties holds:\\
            & \hspace{1cm} (i) $\ell (M)=0$ and $|\mathcal P(M)|=|E|=|\mathcal S(M)|$;\\
            & \hspace{1cm} (ii) $|\mathcal P(M)|=1$;\\
            & \hspace{1cm} (iii) $|\mathcal S(M)|=1$;\\
            & \hspace{1cm} (iv) $r(M)=|E|$;\\
            & \hspace{1cm} (v) $r(M)=0$.\\
            & (b) Else, M is not uniform.\\
\end{tabular}
\newline{}

\section{Some Polynomially Locked Matroids}

Since 2-sums preserved $k$-lockdness for $k\geq 1$ \cite{Chaourar 2008}, $\ell (M(K_4))=4, \ell (W^3)=3, \ell (Q_6)=2, \ell (P_6)=1, \ell (V_8)=5 $, then we can say:
\begin{thm}
If $k\geq 1$ then $\mathcal L_k$ is closed under 2-sums, contains all the excluded minors of 2-sums of uniform matroids and the V\'amos matroid.
\newline In particular, $\mathcal L_1$ is closed under 2-sums and contains all the above matroids.
\end{thm}
It follows that $\mathcal L_1$ contains strictly 2-sums of uniform matroids. 

\section{Conclusion}

We have given a complete description of all facets of the bases polytope of a matroid and deduce two consequences. One about MWBP and the second about TUM.
Future investigations can be characterizing some or all  polynomially locked classes of matroids.

\begin{thebibliography}{1}

\bibitem{Chaourar 2008}
B. Chaourar (2008), {\em On the Kth Best Basis of a Matroid}, Operations Research Letters 36 (2), 239-242.

\bibitem{Chaourar 2011}
B. Chaourar (2011), {\em A Characterization of Uniform Matroids}, ISRN Algebra, Vol. 2011, Article ID 208478, 4 pages, doi:10.5402/2011/208478.

\bibitem{Edmonds 1971}
J. Edmonds (1971), {\em Matroids and the greedy algorithm}, Mathematical Programming 1, 127-136.

\bibitem{Feichtner and Sturmfels 2005}
E. M. Feichtner and B. Sturmfels (2005), {\em Matroid polytopes, nested sets and Bergman fans}, Portugaliae Mathematica 62 (4), 437-468.

\bibitem{Fujishige 1984}
S. Fujishige (1984), {\em A characterization of faces of the base polyhedron associated with a sub modular system}, Journal of the Operations Research Society of Japan 27, 112-128.

\bibitem{Giles 1975}
R. Giles (1975), {\em Submodular functions, Graphs and Integer Polyhedra}, PhD thesis, University of Waterloo.

\bibitem{Jensen and Korte 1982}
P. M. Jensen and B. Korte, {\em Complexity of Matroid Property Algorithms}, SIAM J. COMPUT. 11 (1): 184-190.

\bibitem{Mayhew 2008}
D. Mayhew (2008), {\em Matroid complexity and nonsuccinct descriptions}, SIAM Journal on Discrete Mathematics, 22 (2): 455–466.

\bibitem{Oxley 1992}
J. G. Oxley (1992), {\em Matroid Theory}, Oxford University Press, Oxford.

\bibitem{Pulleyblank 1989}
W. R. Pulleyblank (1989), {\em Polyhedral Combinatorics}, in Handbooks in Operations Research and Management Science (G. L. Nemhauser et al. editors), 371-446.

\bibitem{Robinson and Welsh 1980}
G. C. Robinson and D. J. A. Welsh (1980), {\em The computational complexity of matroid properties}, Math. Proc. Cambridge Phil. Society 87, 29-45.

\bibitem{Schrijver 1986}
A. Schrijver (1986), {\em Theory of Linear and Integer Programming}, John Wiley and Sons, Chichester.







\end{thebibliography}

\end{document}